\documentclass[compress]{elsarticle}
\usepackage{graphicx}
\usepackage{url}
\usepackage{amssymb}
\usepackage{microtype}
\usepackage{inconsolata}
\usepackage{subfig}
\usepackage{listings}
\usepackage{xcolor}
\usepackage{textcomp}
\usepackage{booktabs}

\lstdefinestyle{customc}{%
  belowcaptionskip=1\baselineskip,
  breaklines=true,
  xleftmargin=\parindent,
  language=C,
  showstringspaces=false,
  basicstyle=\scriptsize\ttfamily,
  keywordstyle=\bfseries\color{green!40!black},
  numberstyle=\tiny,
  commentstyle=\itshape\color{purple!40!black},
  identifierstyle=\bfseries\color{black},
  stringstyle=\color{orange},
   morekeywords={uint64_t,uint32_t,__m256i,__m128i,UINT64_C},
}

\usepackage[utf8x]{inputenc}

\usepackage{underscore}

\usepackage{graphicx}

\begin{document}
\begin{frontmatter}

%% Title, authors and addresses
%https://www.elsevier.com/journals/journal-of-computational-and-applied-mathematics/0377-0427/guide-for-authors

%https://en.wikipedia.org/wiki/Xorshift https://lemire.me/blog/2017/09/08/the-xorshift128-random-number-generator-fails-bigcrush/ 

%\title{Reversed Scrambled  Xorshift Generators Xorshift1024*, Xorshift1024+, and Xorshift128+ Fail Big~Crush}
\title{Xorshift1024*, Xorshift1024+, Xorshift128+ and Xoroshiro128+ Fail Statistical Tests for Linearity}

%% use the tnoteref command within \title for footnotes;
%% use the tnotetext command for the associated footnote;
%% use the fnref command within \author or \address for footnotes;
%% use the fntext command for the associated footnote;
%% use the corref command within \author for corresponding author footnotes;
%% use the cortext command for the associated footnote;
%% use the ead command for the email address,
%% and the form \ead[url] for the home page:
%%
%% \title{Title\tnoteref{label1}}
%% \tnotetext[label1]{}

\journal{Journal of Computational and Applied Mathematics}
 \author[UQAM]{Daniel Lemire\corref{cor1}} \ead{lemire@gmail.com} 
 \author[HMC]{Melissa E. O'Neill} \ead{oneill@cs.hmc.edu}  
 
 \address[UQAM]{\scriptsize Universit\'e du Qu\'ebec (TELUQ), 
 5800 Saint-Denis, Montreal, QC, H2S 3L5 Canada
}
\address[HMC]{\scriptsize Harvey Mudd College, 301 Platt Boulevard, Claremont, CA 91711, USA}

 \cortext[cor1]{Corresponding author. Tel.: 00+1+514 843-2015
   ext. 2835; fax: 00+1+800 665-4333.} 

\begin{abstract}
L'Ecuyer \& Simard's \emph{Big~Crush} statistical test suite has revealed statistical flaws in many popular random number generators including Marsaglia's Xorshift generators. 
Vigna recently proposed some 64-bit variations on the Xorshift scheme that are further \emph{scrambled} (i.e., Xorshift1024*, Xorshift1024+, Xorshift128+, Xoroshiro128+). Unlike their unscrambled counterparts, they pass Big~Crush when interleaving blocks of 32~bits for each 64-bit word (most significant, least significant, most significant, least significant, etc.).
We report that these scrambled generators systematically fail Big Crush---specifically the linear-complexity and matrix-rank tests that detect linearity---when taking the 32~lowest-order bits in reverse order from each 64-bit word.

\end{abstract}

\begin{keyword}
Random Number Generator \sep Statistical Tests \sep Xorshift
%% keywords here, in the form: keyword \sep keyword

%% MSC codes here, in the form: \MSC code \sep code
%% or \MSC[2008] code \sep code (2000 is the default)

\end{keyword}

\end{frontmatter}

\section{Introduction}%
\label{section:introduction}%

Pseudorandom number generators (PRNGs) are useful in simulations, games, testing, artificial intelligence, probabilistic algorithms, and so forth. To make it easier to thoroughly test random number generators,
L'Ecuyer and Simard published the 
TestU01 software library~\cite{LEcuyer:2007:TCL:1268776.1268777}. This
freely available and widely used library supports several batteries of tests, the 
most thorough being Big~Crush.

As an instance of particularly fast random number generators, 
Marsaglia proposed the Xorshift family~\cite{marsaglia2003xorshift}.
In C, it can be implemented as a sequence of shift and xor operations 
(e.g., \texttt{xˆ=(x<<13); xˆ=(x>>7); xˆ=(x<<17)}) repeatedly 
applied on an  
integer value initialized from a user-provided seed. At each step,
the returned random value is  the state variable (e.g., \texttt{x}). 
Though  the Xorshift generators are fast,
Panneton and L'Ecuyer showed that they are statistically unreliable~\cite{Panneton:2005:XRN:1113316.1113319}.
In particular, many Xorshift generators fail matrix-rank tests, which  generate random 
$L \times L$~binary matrices, and compare their ranks against the expected theoretical distribution.
Following Marsaglia himself, Panneton and L'Ecuyer proposed 
that the Xorshift generators could be improved by combining them with other operations.

In this spirit, Vigna proposed a 64-bit alternative, xorshift1024* (see Fig.~\ref{fig:xorshift1024star}): he states that it
passes Big~Crush, even after reversing the bit order~\cite{Vigna:2016:EEM:2956571.2845077}. It combines a Xorshift generator with a multiplication.
In a related attempt to improve the statistical properties of Xorshift
generators, Saito and Matsumoto proposed the xorshift-add generator
where two consecutive 32-bit outputs of a Xorshift generator are added together
and returned as the random value~\cite{xorshiftadd2014}. They report
that xorshift-add passes Big~Crush.
However, Vigna observes that if the bit order of the 32-bit random values is reversed (e.g., \texttt{0xff444881}
becomes \texttt{0x811222ff}) before passing the results to Big~Crush, 
the xorshift-add generator systematically fails~\cite{Vigna:2016:EEM:2956571.2845077} both matrix-rank and linear-complexity tests.

In further work, Vigna proposed xorshift128+ and xorshift1024+ (see Figs.~\ref{fig:xorshift1024plus} and~\ref{fig:xorshift128plus}), two new 64-bit
generators that resemble Saito and Matsumoto's xorshift-add in that
 they return the addition of two state values~\cite{Vigna:2017:SMX:3030059.3030179}.
 He again states that they pass Big~Crush. However, unlike
xorshift-add but like xorshift1024*, he adds that they pass Big~Crush even with their bits reversed.
A version of Vigna's xorshift128+ generator has been adopted by the V8
JavaScript engine used by the popular Chrome browser. V8 uses code identical to the
function recommended by Vigna (Fig.~\ref{fig:xorshift128plus}), but where
the constants 23, 18, 5 are replaced by the constants 23, 17, 26.
In more recent work, Blackman and Vigna proposed xoroshiro128+ (see Fig.~\ref{fig:xoroshiro128}),
presented as the successor of xorshift128+~\cite{xoroshiro128plus}.
%http://vigna.di.unimi.it/xorshift/xoroshiro128plus.c
We refer to these generators (xorshift1024*, xorshift1024+, xorshift128+ and 
xoroshiro128+) as \emph{scrambled Xorshift} generators, following
Vigna's terminology.

These scrambled Xorshift generators output
64-bit integers. Ideally, a 64-bit PRNG should be tested with a 64-bit test suite; using a test suite designed for 32-bit PRNGs (or actually 31-bit PRNGs in the case of Big~Crush) necessarily involves compomises.  Vigna's approach was to interleave 
the least-significant 32~bits and
most-significant 32~bits of each 64-bit output, thus each 64-bit output becomes two 32-bit outputs.  This interleaving strategy may detect flaws in some 64-bit PRNGs, but it may also hide statistical weaknesses. Such an approach is useful but is \emph{insufficient}.  A more comprehensive strategy is to also test the most-significant and least-significant 32~bits separately.
When we focus on the least-significant 32~bits, and, specifically,
in the case where their bits are reversed, 
we find that the scrambled Xorshift generators systematically
fail Big~Crush. 

 \begin{figure}[htb]
     \centering     \subfloat[xorshift1024*\label{fig:xorshift1024star}]{\begin{tabular}{c} 
\begin{lstlisting}
uint64_t s[16];
int p;
uint64_t mult = 0x106689D45497FDB5;
uint64_t xorshift1024star(void) {
  uint64_t s0 = s[p];
  uint64_t s1 = s[p = (p + 1) & 15];
  s1 ^= s1 << 31; 
  s[p] = s1 ^ s0 ^ (s1 >> 11) 
        ^ (s0 >> 30); 
  return s[p] * mult;
}
\end{lstlisting}
\end{tabular}
}
     \subfloat[xorshift1024+\label{fig:xorshift1024plus}]{\begin{tabular}{c} 
\begin{lstlisting}
uint64_t s[16];
int p;
uint64_t xorshift1024plus(void) {
  uint64_t s0 = s[p];
  uint64_t s1 = s[p = (p + 1) & 15];
  uint64_t result = s0 + s1;
  s1 ^= s1 << 31; 
  s[p] = s1 ^ s0 
        ^ (s1 >> 11) 
        ^ (s0 >> 30); 
  return result;
}
\end{lstlisting}
\end{tabular}}

    \subfloat[xorshift128+\label{fig:xorshift128plus}]{\begin{tabular}{c} 
\begin{lstlisting}
uint64_t s[2];
uint64_t xorshift128plus(void) {
  uint64_t s1 = s[0];
  uint64_t s0 = s[1];
  uint64_t result = s0 + s1;
  s[0] = s0;
  s1 ^= s1 << 23;
  s[1] = s1 ^ s0 
    ^ (s1 >> 18) ^ (s0 >> 5); 
  return result;
}
\end{lstlisting}\end{tabular}}
\subfloat[xoroshiro128\label{fig:xoroshiro128}]{\begin{tabular}{c} 
\begin{lstlisting}
uint64_t s[2];
uint64_t rotl(uint64_t x, int k) {
 return (x << k) | (x >> (64 - k));
}
uint64_t xoroshiro128(void) {
  uint64_t s0 = s[0];
  uint64_t s1 = s[1];
  uint64_t result = s0 + s1;
  s1 ^= s0;
  s[0] = rotl(s0, 55) ^ s1 ^ (s1 << 14);
  s[1] = rotl(s1, 36);
  return result;
}
\end{lstlisting}\end{tabular}}

\caption{C functions defining the various scrambled Xorshift generators\label{fig:ccode}}
\end{figure}

\section{Results}

It is expected that even good generators could
fail some tests, some of the time.
We are only interested in decisive and systematic failures:
\begin{itemize}
\item We focus solely on tests failed
with extreme $p$-values according to TestU01. By excluding mild failures (e.g., $p = 0.05$), we reduce 
the likelihood of false negatives. 
\item We only report failures that occur irrespective of the initial seed. 
\end{itemize}

Thus we test the scrambled xorshift generators
with one hundred different initial seeds 
(denoted by the variable \texttt{s} in Fig.~\ref{fig:ccode}).
The generators under consideration  require
relatively long seeds (at least 16~bytes).
To generate a sufficiently long seed (128 or 1024~bits),
we produce a random integer with the Bash shell's internal RANDOM function 
as an initial seed for to the 64-bit SplitMix generator~\cite{Steele:2014:FSP:2714064.2660195}. 
The SplitMix generator is called twice for the generators requiring 128-bit seeds
and sixteen~times for the generators requiring 1024-bit seeds.
According to our tests, the 64-bit SplitMix generator
passes Big~Crush, even with the bit order reversed.

We use the latest version of TestU01 (version 1.2.3).
We report the tests failed in Big~Crush for each generator in Table~\ref{tab:failed};
we only report tests that failed for all one hundred different seeds. All
scrambled generators fail the linear-complexity tests (LinearComp). The 128-bit xorshift+ generators additionally fail the matrix-rank tests (MatrixRank) for both $L=1000$ and $L=5000$.
All matrix-rank tests  fail with a $p$-value smaller than $ 10^{-100}$,
all linear-complexity tests fail with a $p$-value greater than or equal to $1-10^{-15}$,
except for one seeding of xorshift128+, where we observed a $p$-value of $1-4.3 \times 10^{-11}$. Getting such extreme $p$-values for one hundred different seeds indicates a systematic failure.
There are many more minor, nonsystematic
failures (e.g., 33 for xorshift128+) that might require further analysis~\cite{Haramoto2009}.
To ease reproducibility, both our scripts and results are freely  available online.\footnote{See \url{https://github.com/lemire/testingRNG}.}

\begin{table}
\centering\small
\begin{tabular}{lp{5cm}}
\toprule
Scrambled 64-bit XorShift & Failed 32-bit Big~Crush tests\\
\midrule
xorshift1024* & LinearComp \\
xorshift1024+ & LinearComp \\
xorshift128+ & LinearComp, MatrixRank  \\
xorshift128+ (v8) & LinearComp, MatrixRank    \\
xoroshiro128 & LinearComp, MatrixRank \\
\bottomrule
\end{tabular}
\caption{Tests failed systematically by the least significant 32~bits in reversed bit order\label{tab:failed}. }
\end{table}

\section{Conclusion}

The least significant 32~bits of the 64-bit scrambled xorshift generators
(xorshift1024*, xorshift1024+, xorshift128+ and 
xoroshiro128+) systematically fail Big~Crush. In particular,  
all fail the linear-complexity tests.
These scrambled xorshift generators are derived from 
Marsaglia's xorshift generators which also systematically fail these tests~\cite{LEcuyer:2007:TCL:1268776.1268777}.

Panneton et al.~\cite{Panneton:2006:ILG:1132973.1132974} note that PRNG schemes based  on linear recurrences modulo~2---which includes xorshift schemes with simple output functions---can expect to see detectable linear dependencies in tests such as the matrix-rank test~\cite{rMAR85a} and the linear-complexity tests~\cite{rERD92a}, so the failures we report are unsurprising. Panneton et al.\ also note that although these problems are an issue for some applications, for many applications they would go  unnoticed. Thus readers should not assume that failing some statistical tests renders a PRNG scheme  worthless.  But it is useful to note that there are multiple PRNG schemes available that do not fail any currently available statistical tests and thus do not require weighing whether detectable linearity is a problem for a particular use case or not~\cite{LEcuyer:2007:TCL:1268776.1268777}. % ~\cite{Steele:2014:FSP:2714064.2660195}.

Mirroring a related recommendation by Press et al.~\cite{Press:2007:NRE:1403886}, Vigna argued that  
TestU01 should always be applied on the reversed generators~\cite{Vigna:2016:EEM:2956571.2845077} to address the problem of linearity issues in the low-order bits of a PRNG going undetected. Moreover, our own results
suggest that when assessing 
a 64-bit generator with TestU01, thorough testing requires us to  
test the least significant 32~bits separately
from the most significant 32~bits, otherwise statistical issues issues may likewise go undetected.

%\setlength{\bibsep}{0pt plus 0.3ex}
%(Verify that the bibliography style matches the journal)
\bibliographystyle{model1-num-names}
\bibliography{scrambledfail.bib}

\end{document}